%% file: main.tex
\documentclass[10pt,journal,compsoc]{IEEEtran}
\ifCLASSINFOpdf
\else
\fi
\usepackage[space,nocompress]{cite}
\usepackage{graphicx,caption,amssymb,amsmath,array}
\usepackage{color,float}
\usepackage{subfigure,epsfig}
\usepackage{ragged2e}
\usepackage{enumitem}
\usepackage{footnote}

\usepackage{algorithm}
\usepackage{algorithmic}
\usepackage{multirow}
\usepackage{float}
\usepackage{url}
\usepackage{enumitem}
\usepackage{array}
\usepackage{booktabs}
\usepackage{bm}
\usepackage{color}

\setlist[itemize]{leftmargin=*}
\setlist[enumerate]{leftmargin=*}
\begin{document}
\title{Implicit semantic-based personalized micro-videos recommendation}

\author{Bo~Liu~\IEEEmembership{, Member, IEEE}
}

\markboth{Journal of \LaTeX\ Class Files,~Vol.~14, No.~8, August~2015}%
{Shell \MakeLowercase{\textit{et al.}}: Bare Demo of IEEEtran.cls for IEEE Transactions on Magnetics Journals}



% for Transactions on Magnetics papers, we must declare the abstract and
% index terms PRIOR to the title within the \IEEEtitleabstractindextext
% IEEEtran command as these need to go into the title area created by
\IEEEtitleabstractindextext{%
\justifying  
\begin{abstract}
With the rapid development of mobile Internet and big data, a huge amount of data is generated in the network, but the data that users are really interested in a very small portion. To extract the information that users are interested in from the huge amount of data, the information overload problem needs to be solved.In the era of mobile internet, the user's characteristics and other information should be combined in the massive amount of data to quickly and accurately recommend content to the user, as far as possible to meet the user's personalized needs. Therefore, there is an urgent need to realize high-speed and effective retrieval in tens of thousands of micro-videos. Video data content contains complex meanings, and there are intrinsic connections between video data. For multimodal information, subspace coding learning is introduced to build a coding network from public potential representations to multimodal feature information, taking into account the consistency and complementarity information under each modality to obtain a public representation of the complete eigenvalue. An end-to-end reordering model based on deep learning and attention mechanism, called interest-related product similarity model based on multimodal data, is proposed for providing top-N recommendations.The multimodal feature learning module, interest-related network module and product similarity recommendation module together form the new model.By conducting extensive experiments on publicly accessible datasets, the results demonstrate the state-of-the-art performance of our proposed algorithm and its effectiveness.
\end{abstract}

}

\maketitle

\IEEEdisplaynontitleabstractindextext

% \fi
\IEEEpeerreviewmaketitle

% \input{1_introduction.tex}
% \input{2_related.tex}
% % \input{Data.tex}
% \input{3_method.tex}
% \input{4_experiment.tex}
% \input{5_conclusion.tex}
\input{1_introduction.tex}

\input{2_preliminary.tex}
\input{3_method.tex}

\input{4_experiment.tex}
\input{5_conclusion.tex}
\bibliography{BIB/IEEEabrv, reference}

\end{document}

%% file: 1_introduction.tex
% \newpage
\section{Introduction}
With the rapid development of mobile Internet and big data, a huge amount of data is generated in the network, but the data that users are really interested in a very small portion. To extract the information that users are interested in from the huge amount of data, the information overload problem needs to be solved. Baidu and Google use information retrieval and search engine technology to locate data of interest to users, which to some extent simply solves the information overload problem. However, these search engines do not consider the characteristics of different users, and when users search for the same keywords, the system retrieves the same content, and there is a large amount of redundant information of no value to users, which cannot meet the personalized needs of users. In the era of mobile internet, the user's characteristics and other information should be combined in the massive amount of data to quickly and accurately recommend content to the user, as far as possible to meet the user's personalized needs. Therefore, there is an urgent need to realize high-speed and effective retrieval in tens of thousands of micro-videos. Video data content contains complex meanings, and there are intrinsic connections between video data. In visual modality, association relationships exist between visual features and features, between video clips and clips, and between video semantics and semantics. Through these correlation relationships, the computational cost of semantic detection can be reduced and the quality of search can be improved. Correlations between video semantics have an important role, and problems such as synonymy and polysemy can occur between video semantic contents, which are caused by ignoring the correlations of video semantics and can lead to a series of defects.

In recent years, a great deal of progress has been made in the study of problems related to preference and click-through prediction~\cite{SI2}. The most widely used prediction method in industry is the use of logistic regression (LR) to learn click-through prediction models~\cite{chapelle2014simple,richardson2007predicting,mcmahan2013ad}.LR has the advantage of being simple and very easy to implement for massively real-time parallel processing, but linear models have limited learning ability and cannot capture the information carried by higher-order features (nonlinear information) ~\cite{zhang2016deep}. Joachims et al.~\cite{joachims2002optimizing} proposed a SupportVector Machine (SVM) model to predict ad click-through rates, which can effectively handle multidimensional nonlinear data but cannot predict sparse ads with large data volumes.Lee et al.~\cite{lee2018estimating} modeled the data hierarchically from a tripartite perspective of media, users, and advertisers. Shen et al.~\cite{shen2012personalized} proposed a click-through rate prediction model based on collaborative filtering and tensor decomposition. The model mines users' personalized preferences based on the relationship between users, queries, and documents to improve prediction accuracy.Rendle et al.~\cite{BPR} combined the advantages of support vector machines (SVM) and decomposition models to propose a Factorization Machines (FM) model, which uses decomposition parameters to model all interactions between variables and can be used in very sparse data for parameter estimation and has better prediction quality compared to SVM, moreover, FM is a general-purpose predictor that can be used with any real-valued feature vector.

The current research on micro-videos is mainly focused on the following directions. Micro-video scene classification: e.g., Liu et al.~\cite{liu2017towards} capture the short duration of micro-videos and propose an end-to-end model that jointly models sparsity and multi-sequence structure as a way to capture the temporal structure between video frames and the sparse representation of Micro-videos; Micro-video popularity prediction: Jing et al.~\cite{jing2017low}  propose a direct push learning method based on low-rank multi-view embedding to seek a set of specific low-rank constrained projection matrices of views to map multi-view features into a common subspace; in addition, graph regularization terms are constructed to improve generalization ability and further prevent overfitting problems.Micro-video user recommendation: Chen et al.~\cite{chen2020learning} proposed a deep neural network model that fuses multiple interest feature representations of users, and used a fusion of four user interest representations: user information, entry information, history information and neighbor information in the paper. For micro-videos as multimedia contents often have multiple labeled information, the author works on the problem of micro-video multi-label classification. The traditional supervised learning is to learn the mapping relationship from sample space $X$ to label space $Y$, i.e., W: $X\rightarrow Y$~\cite{yu2020personalized}. At this time, $Y$ is mostly a label with unique semantic information, unlike this, in reality, there are co-occurrence properties of each label, for example, "singing" and "playing guitar often appear in the same sample at the same time, and there is a strong semantic correlation between the tags, so the task becomes to learn the mapping relationship from the sample space to the tag space with multiple semantic information, and this relationship is one-to-many. Existing approaches to multi-label learning are as follows: converting multi-label classification problems to binary or multi-classification problems in order to utilize classification algorithms that achieve good results on traditional supervised learning; using low-rank representations to obtain a common low-rank representation, such as Jia et al.~\cite{jia2019facial} proposed a method that uses low-rank structure learning model to stably obtain local correlations between labels; Chen et al.~\cite{d2015review} proposed a graph-based semantic association regularization method for enhancing representation learning, which is mainly used to explore interconnections in terms of personal attributes as a way to perform privacy detection of UGCs. Based on the above analysis, this paper focuses on the research and implementation of micro-video personalized recommendation algorithm with multimodal data sources:
\begin{itemize}
    \item For multimodal information, subspace coding learning is introduced to build a coding network from public potential representations to multimodal feature information, taking into account the consistency and complementarity information under each modality to obtain a public representation of the complete eigenvalue. 
    \item An end-to-end reordering model based on deep learning and attention mechanism, called interest-related product similarity model based on multimodal data, is proposed for providing top-N recommendations.The multimodal feature learning module, interest-related network module and product similarity recommendation module together form the new model.
    \item By conducting extensive experiments on publicly accessible datasets, the results demonstrate the state-of-the-art performance of our proposed algorithm and its effectiveness.
\end{itemize}

\begin{figure}
	\centering
    \includegraphics[width=0.45\textwidth]{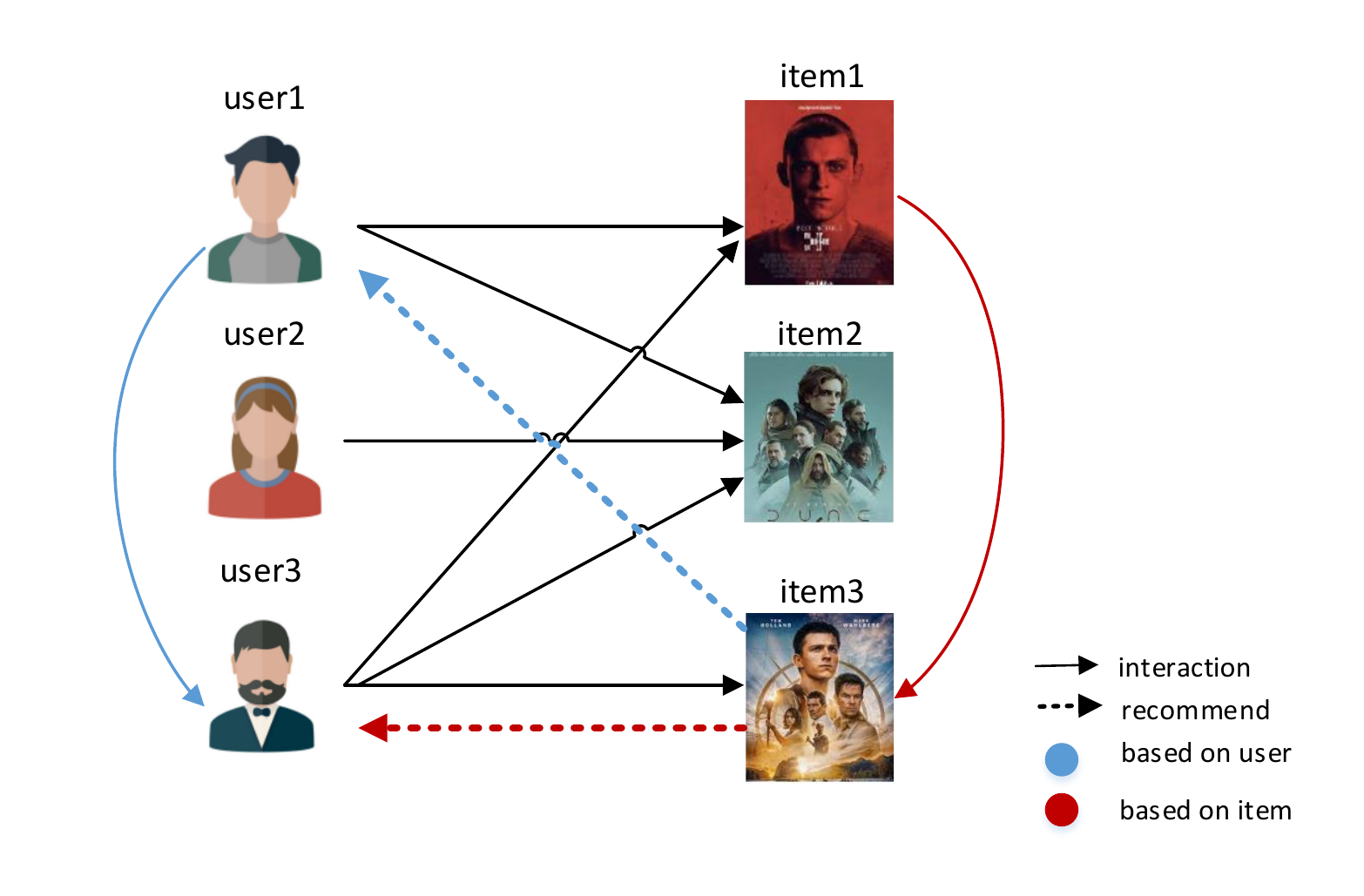}
    % \vspace{-5pt}
    \caption{An illustration of CF-based recommendation. }
    % \vspace{-15pt}
	\label{fig:figure1}
\end{figure}

%% file: 2_preliminary.tex
\section{Related Work}
The study of video recommendation systems has affected all aspects of science, society, economy, and life, and they are gradually becoming one of the most powerful and popular tools nowadays as tools that can generate huge business benefits~\cite{KG}. Personalized recommendation system is an intelligent algorithm and decision based on big data. The online recommendation system makes personalized matching by capturing users' historical behavior data (browsing, viewing, purchasing, rating, etc.) and the content information of the sold products (profile, details, applicable scenarios, etc.), and corrects and optimizes its recommendation results based on the feedback results of interaction with users (clicks, ratings, etc.)~\cite{LightGCN,wei2021contrastive}.
\subsection{Collaborative Filtering}

In 1992, Tapestry, the first system using Collaborative Filtering (CF), emerged to predict user preferences by learning from user-product interactions, and soon attracted strong interest in academia and industry because of its data availability and simplicity. The core idea of CF is to generate effective recommendations based on historical user interactions by mining the implied user relationships, product relationships, and user-product relationships. The data collection relies on the back-end system to track users' online footprints, such as browsing, clicking, rating, purchasing, viewing, etc. Specifically, CF~\cite{suganeshwari2016survey} can be divided into memory-based collaborative filtering and model-based collaborative filtering.User-based CF identifies users whose preferences for products are close to those of the users to be predicted, and thus makes recommendations for the users to be predicted based on the preferences of this group of users for new products.Item-based CF is more commonly used, and this algorithm was first proposed by Amazon and put into use~\cite{linden2003amazon}. As in Figure 1, Item-based CF aims to find products that are similar to the new product and use the user base that prefers similar products as the deliverable user base for the new product. In contrast, model-based CF is based on historical rating data and uses machine learning models to mine correlations and make predictions.Simon Funk improved on the traditional SVD by proposing the Funk-SVD algorithm, which represents products and users in the same hidden space to reflect the hidden correlations between users and products. Later, Yehuda Koren trained SVD based on gradient descent method in Netflix algorithm competition and named it as hidden semantic model~\cite{bell2007modeling}. Koren improved and perfected SVD based on SVD to make it applicable to hidden feedback scenario that proposed SVD++ algorithm~\cite{koren2008factorization}. The later proposed TimeSVD++~\cite{koren2009collaborative} considers that the longer the preference is from the current prediction, the less reliable it is, and the closer the preference is to the current prediction, the more influential it is. Factorization Machines (FM) as well as extension algorithms were proposed and continuously improved the recommendation performance~\cite{MF}.

\subsection{Content-based Recommendations}

Unlike collaborative filtering, which requires content information about the item (e.g., title,  description) to compare the similarity between items and does not take into account the historical relevance between users and items. The descriptive information of the product is processed by data as a feature vector thus used to create a model reflecting the user's preferences. Content-based recommendation methods utilize information containing user tastes, preferences, and needs as well as a variety of models modeled for specific purposes, where the modeling information can come from explicit or implicit feedback from users~\cite{salton1975vector,jiang2020aspect}, and the available modeling models are various machine learning models, vector space models, etc.Content-based filtering (CBF) models require as input the target user’s behavioral data (but not those of non-target users’) together with item content information, typically represented as real-values vectors. To give some examples in the multimedia domain, item properties can be words or concepts in a text, colors in an image, amount of motion in a movie, or rhythm in a music piece.

\subsection{Deep Learning-based Recommendations}
Due to the excellent performance capabilities of deep learning on many complex tasks, academia and industry have competed to extend it to a wider range of applications and achieve state-of-the-art results so far. Recommender systems are an important part of the industrial landscape. For many online websites and mobile applications, it is an important tool for promoting sales and services. Recently, many researchers have adopted deep learning to further improve the quality of their recommendations~\cite{covington2016deep,KGAT,MMGCN,cheng2016wide,tao2020mgat}.Covington et al.~\cite{covington2016deep} implemented a recommendation algorithm based on a multilayer perceptron on the YouTube platform with 1.9 billion users. Cheng et al.~\cite{cheng2016wide} proposed an app recommendation system for Google Play called Wide\&Deep, which uses a hybrid model of linear and deep models.The news recommendation business in the Yahoo website uses an RNN-based news recommendation system~\cite{okura2017embedding}.Application of MLP in recommendation algorithms:MLP is concise and clear in terms of mathematical theory and has been shown to approximate any network measurable function to any desired accuracy~\cite{hornik1989multilayer}.NCF~\cite{NCF} is a framework that captures both linear and nonlinear relationships between users and products.ACF ~\cite{ouyang2014autoencoder} is the first collaborative recommendation model based on autoencoders. Instead of using the original partial observation vector, it is decomposed into integer ratings. CFN~\cite{strub2016hybrid} takes a vector of users or products with auxiliary information such as user attributes and product descriptions as input then reconstructs it at the output layer, where the loss function is the gap between input and output. 
Application of CNNs in recommendation algorithms: CNNs are very effective for processing visual, textual and speech information, and most of them are used to extract features in recommendations. Wang et al.~\cite{wang2017your} used CNNs to extract image features and proposed a visual content-enhanced point-of-interest recommendation system called VPOI, which confirmed the key role of visual features for point-of-interest recommendations. ConTagNet~\cite{rawat2016contagnet} is a tag recommendation system that introduces context-awareness, where one module uses a neural network to learn image features and another module uses a two-layer fully connected feed forward neural network to model the contextual representation, and finally the outputs of the two neural networks are connected into a softmax function thus predicting the probability of each candidate tag. Hidasi et al.~\cite{hidasi2015session} proposed a GRU-based session recommendation system. The input is the product coding status associated with the session say N products, then 1 at the products associated with the session and 0 at the irrelevant products, then all the products are coded as a vector of length N. The output is the maximum likelihood probability that each product may be the next relevant product.RRN~\cite{wu2017recurrent} is a nonparametric recommendation model built on RNNs that can model the seasonal evolution of products and changes in user preferences over time, utilizing two LSTM networks to model dynamic user states and product states, and to account for fixed long-term user interests and static product characteristics, the model also integrates the hidden factor vectors of users and items.

%% file: 3_method.tex
\section{Methodology}

\subsection{Model Framework}

Given a set of micro-video datasets $\mathcal{X}=\{X^{(1)},X^{(2)},\dots,X^{(V)}\}$,where $V$ is the number of views, $X^{(V)}\in {R^{D_v \times N}}$ is the feature matrix under the v-th view, $D_v$is the feature dimension under the v-th view, $N$ is the number of short video samples. Also given the label matrix$Y=Y_{ij}\in \{0,1\}^{N\times C}$, where $C$ is the number of labels. If the i-th micro-video carries the j-th label, then the corresponding $Y_{ij}=1$ , and vice versa, $Y_{ij}=0$.

Assuming the existence of a common subspace that can portray features under different perspectives, with the aim of making full use of information from each perspective.Suppose there is a shared common representation $H$ , which can reconstruct the features of each sample at each viewpoint by a set of mappings, i.e., $X^{(V)}=f_v(H)$, where, for any viewpoint $v$, construct the reconstruction mapping$f_v(\cdot)$ belonging to its corresponding viewpoint. In the following, the reconstructed mappings of sample features are constructed using multilayer perceptual neural networks for each of the three modalities; in addition, for the same sample, different modalities share the same complete common representation $H$ , which encodes the sample features $X^{(V)}$ in multiple modalities.

Given a common representation each perspective is conditionally independent. The key to learning for a multi-view subspace is to effectively encode the information for the different available perspectives as:
\begin{equation}
p(\mathcal{X}|H)=p(X^{(1)}|H)p(X^{(2)}|H)\dots p(X^{(v)}|H),
\end{equation}

For the given $\mathcal{X}$, the likelihood modeling of the common representation$H$ can be represented as:
\begin{equation}
p(\mathcal{X}|H)\propto e^{-\Delta (x,f_v(H))},
\end{equation}
where,$H=\{h_n\}_{n-1}^N\in R^{D\times N}$,$h_n$ corresponding to the embedding representation of n-th sample.It can be found that maximizing the likelihood function is equivalent to minimizing the reconfiguration loss:
\begin{equation}
min{\Delta (x,f_v(H))}=min\sum _{v-1}^v\lvert| U^{(v)}-X^{(v)}\rvert|^2,
\end{equation}
where $X^{(v)}$ is the sample feature matrix at the v-th viewpoint and $U^{(v)}$ is the output of the sample at the v-th viewpoint after reconstruction mapping. The reconstructed loss from input to output of the sample is established and the reconstructed network is trained with its constraints. Its column vector $u^{(v)}_n$ is specified as:
\begin{equation}
u^{(v)}_n=\theta(W^{(v)}_h h_n +b^{(v)}_h),
\end{equation}
where $W^{(v)}_h$ and $b^{(v)}_h$ denote the weight and bias of the network at the v-th viewpoint, respectively. $\theta(\cdot)$ is the activation function.

\begin{figure*}
	\centering
    \includegraphics[width=0.9\textwidth]{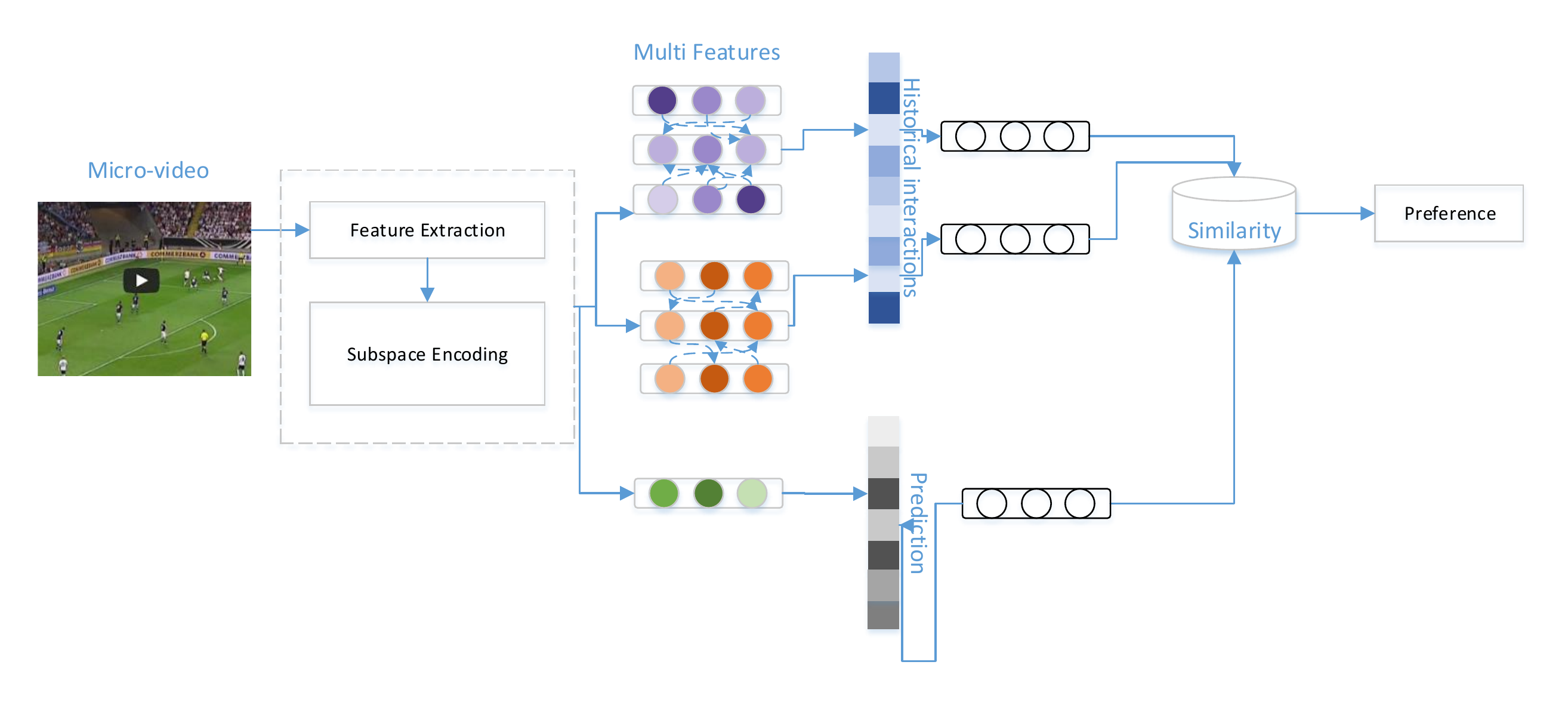}
    % \vspace{-5pt}
    \caption{An illustration of our model framework.}
    % \vspace{-15pt}
	\label{fig:figure1}
\end{figure*}

\subsection{Feature Filtering}
Suppose the number of scene categories is $N_{scene}$, for the i-th micro-video, the purpose of scene recognition is to find the the maximum value of the scene prediction probability:
\begin{equation}
p_{v_i}^s=max(p_{v_i}^{s_j}) (1\leqslant j \leqslant N_{scene},1\leqslant i \leqslant n),
\end{equation}
where $v_i$ is the i-th micro-video, and $p_{v_i}^{s_j}$ is the probability value of the i-th micro-video corresponding to the j-th scene. However, it is necessary to keep the probability values of $v_i$ in all scenes to retain as much useful information in the video as possible information:
\begin{equation}
f_{v_i}^s=\{p_{v_i}^{s_j}\} (1\leqslant j \leqslant N_{scene},1\leqslant i \leqslant n),
\end{equation}

Next, behavioral features need to be extracted from the micro-video, using variable convolutional network of dual-flow CNN algorithm for RGB feature and Flow feature extraction.  The advantage of the variable convolutional network is that it can change the shape of the convolutional kernel by adding offsets to the convolutional kernel, thus improving the convolutional network's ability to adapt to the image adaptation capability.

Assuming that the number of behavior categories is $N_action$, for the i-th micro-video, the result of behavior recognition can be defined as:
\begin{equation}
p_{v_i}^A=max(p_{v_i}^{a_k}) (1\leqslant k \leqslant N_{action},1\leqslant i \leqslant n),
\end{equation}
where $v_i$ is the i-th micro-video and $p_{v_i}^{a_k}$ is the probability value of the i-th micro-video corresponding to the i-th probability value of the k-th behavior, which can be defined as:
\begin{equation}
p_{v_i}^{a_k}=p_{v_i}^{a_k^{RGB}}p_{v_i}^{a_k^{Flow}},
\end{equation}

Similarly, the behavioral feature extraction part needs to keep the probability values of each micro-video for all behaviors:
\begin{equation}
f_{v_i}^A=\{p_{v_i}^{a_k}\} (1\leqslant k \leqslant N_{action},1\leqslant i \leqslant n),
\end{equation}

According to the above analysis, for the micro-video $V = \{v_1,v_2,\dots,v_n\}$, the joint features are:
\begin{equation}
F_V=\{f_{v_1},f_{v_2},\dots,f_{v_n},\},
\end{equation}

\subsection{Preference Prediction}
After merging the high-impact features extracted by GBDT with discrete features for unique thermal coding, the input feature vector is decomposed into two-by-two factors, and the preference rate of micro-videos is used as the output, and the micro-video preference rate prediction model is defined as:
\begin{equation}
y(x)=w_0 +\sum w_i x_i +\sum_{i=1}^n \sum_{j=i+1}^n (V_i, V_j)x_i x_j,
\end{equation}
where $x_i$ is the value of the i-th feature, $n$ is the dimension of the micro-video feature, and $w_0\in R$is the global bias, $w_i \in R^n$ is the influence factor of the i-th feature, and $V \in  R^{n\times  h}$ is the interaction parameter between the mutually exclusive feature components. $V_i,V_j$ represent the dot product of two vectors $V_i$ and vector $V_j$ of dimension $h$.
\begin{equation}
(V_i, V_j)=\sum _{f=1}^h v_i \cdot V_{(i,f)},
\end{equation}
where $V_i$ denotes the i-th dimensional vector of the coefficient matrix $V$ and $V_i =(v_{i,1},v_{i,2},\dots,v_{i,h}),h\in N ^0$ is the hyperparameter.

%% file: 4_experiment.tex
\section{EXPERIMENTS}

 \begin{table}
  \centering
  
  \renewcommand\arraystretch{1.4}
  \caption{ Statistics of the evaluation dataset.}
    %\vspace{0.1mm}
  \label{table_1}
  \setlength{\tabcolsep}{2.0mm}
  \begin{tabular}{|c|c|c|c|c|}
    \hline
    \textbf{Dataset}&\textbf{Users}&\textbf{Items}&\textbf{Interactions}&\textbf{Density}\\
    \hline
    \textbf{MovieLens}&$6040$&$3685$&$998034$&$4.48\%$\\
    \hline
    \textbf{Amazon}&$39385$&$23033$&$278677$&$0.031\%$\\
    \hline
      \end{tabular}
  \vspace{-2mm}
\end{table}

\subsection{Dataset}
We validated the proposed algorithm on the public dataset MovieLens 1M and a partial subset of the Amazon product dataset Clothing\&Shoes\&Jewelry. Table1 summarizes the statistical properties of the two datasets.

MovieLens: Because standard MovieLens dataset does not contain multimodal information, we used the OMDb API to crawl each movie's corresponding poster and text synopsis as visual information and textual information for validating Multimodal IRIS. We used the presence or absence of historical user ratings as implicit feedback.

Amazon: Amazon includes text and image information for each product. In order to quickly validate our model, we ignore users with less than five historical interactions and remove products with ratings below 5 in preprocessing. We use users' review histories as implicit feedback.

\subsection{Baseline}
VBPR~\cite{he2016vbpr}: VBPR integrates visual information into the prediction of people's preferences. VBPR has a significant improvement over matrix decomposition models that rely only on user hidden vectors and product hidden vectors. We have adapted VBPR to be suitable for learning with logarithmic loss functions.

FISM~\cite{kabbur2013fism}: This model using a mixture of hidden factor models and nearest neighbor based models is consistent with our proposed Multimodal IRIS, and thus the method is a very important benchmarking approach.

NCF~\cite{he2017neural}: NCF is based on a matrix decomposition framework that utilizes MLP to model the nonlinear relationship between users and products. The model achieves the best performance among hidden factor models due to the powerful representation capability of neural networks.

NAIS~\cite{he2018nais}: NAIS is designed to differentiate the importance of different historical products in user modeling. The product similarity recommendation using the attention mechanism is also part of Multimodal IRIS.

\subsection{Evaluation Metrics}
Using all interaction data as positive samples, for each user we randomly select one sample for composing the test set and the rest for the training set. Then we draw K negative samples from the uninteracted products with equal probability for training, K being the hyperparameter used to control the sampling. In addition, the validation phase is not possible for us to sort all the uninteracted products in the validation phase because the set containing all the products is relatively large which leads to a large time consumption. Therefore, for each positive example in the test set, we randomly select 99 negative examples to form a test pair, so that each user in the test set corresponds to a set of 100 candidates, as is common in most research work.

The commonly used measures in typical top-N recommendations based on implicit feedback are Hit Ratio (HR) and Normalized Discounted Cumulative Gain (NDCG). The HR@N is used to measure whether the positive examples in the test set appear in the top-N recommendation list; NDCG@N also considers the position of the positive examples in the test set in the top-N recommendation list, where N is a hyperparameter. We compared the experimental results based on comparing at different N. For both metrics, higher values represent better model performance. The specific calculation is as follows:
\begin{equation}
HR@N=\frac{\sum_{i=1}^N co(i)}{N} ,
\end{equation}
where $co(i)$ indicates whether the positive example retained in the test set appears in the top-N recommendation list of user $i$. If it does, the value is 1, otherwise the value is 0. $N$ indicates the number of users.
\begin{equation}
NDCG@N=\frac{1}{N} \sum_{i=1}^N hits(i),
\end{equation}

   \begin{table}
  \centering
  \renewcommand\arraystretch{1.5}
  \caption{ Experimental results on Movielens dataset.}
    %\vspace{0.1mm}
  \label{table_2}
  \setlength{\tabcolsep}{2.0mm}
  \begin{tabular}{|c|c|c|}
    \hline
    \textbf{Model}&\textbf{HR@10}&\textbf{NDCG@10}\\
    \hline
    \textbf{VBPR}&$0.7345$&$0.5187$\\
    \hline
    \textbf{FISM}&$0.8117$&$0.5753$\\
    \hline
     \textbf{NCF}&$0.8112$&$0.5637$\\
    \hline
     \textbf{NAIS}&$0.8129$&$0.5834$\\
    \hline
     \textbf{Ours}&$0.8287$&$0.5899$\\
    \hline
      \end{tabular}
  \vspace{-2mm}
\end{table}

 \begin{table}
  \centering
  \renewcommand\arraystretch{1.5}
  \caption{ Experimental results on Amazon dataset.}
    %\vspace{0.1mm}
  \label{table_2}
  \setlength{\tabcolsep}{2.0mm}
  \begin{tabular}{|c|c|c|}
    \hline
    \textbf{Model}&\textbf{HR@10}&\textbf{NDCG@10}\\
    \hline
    \textbf{VBPR}&$0.3376$&$0.2087$\\
    \hline
    \textbf{FISM}&$0.4007$&$0.2456$\\
    \hline
     \textbf{NCF}&$0.3654$&$0.2143$\\
    \hline
     \textbf{NAIS}&$0.4233$&$0.2431$\\
    \hline
     \textbf{Ours}&$0.4643$&$0.2879$\\
    \hline
      \end{tabular}
  \vspace{-2mm}
\end{table}

\subsection{Results Analysis}
From these two tables it can be shown that our proposed model outperforms the deep model (VBPR and NCF) and the shallow model (FISM and NAIS). Using only image information as input, the proposed Image IRIS consistently outperforms compared to VBPR on MovieLens and Amazon, and these results indicate that Image IRIS makes fuller use of image information and helps in recommendation. NCF improves the recommendation performance by fusing MF and MLP to learn linear and nonlinear interaction functions between users and products. The improvement of Multimodal IRIS over NCF confirms the effectiveness of multimodal features and neighborhood-based models. Compared to NAIS on the MovieLens dataset, our model improves $0.78\%$ on HR@10, $1.66\%$ on NDCG@10. Also, on the Amazon dataset, our model improves $4.52\%$ on HR@10, $3.87\%$ on NDCG@10. Therefore, we can conclude that the use of multimodal information can improve the recommendation efficiency. It is worth pointing out that the difference in performance between FISM and NAIS on the Amazon dataset is very small (Amazon is two orders of magnitude sparser than MovieLens). Due to the extreme sparsity of the data, it is not possible to efficiently learn the attention model in NAIS by hidden factors alone. Our model produced more significant results on the Amazon dataset compared to MovieLens, which revealed the necessity of adding multimodal information when dealing with sparse datasets. By comparing Image-add-Text IRIS and Our model, we can see that the use of knowledge sharing units in Our model further improves the evaluation metric values compared to the traditional way of using multimodal information, i.e., direct weighted summation of multimodal data, indicating that knowledge transfer helps feature learning. In conclusion, our model provides superior performance compared to existing methods.

\begin{figure}
    \centering
    \subfigure[NDCG@N on Movielens]{
      \includegraphics[width=0.23\textwidth]{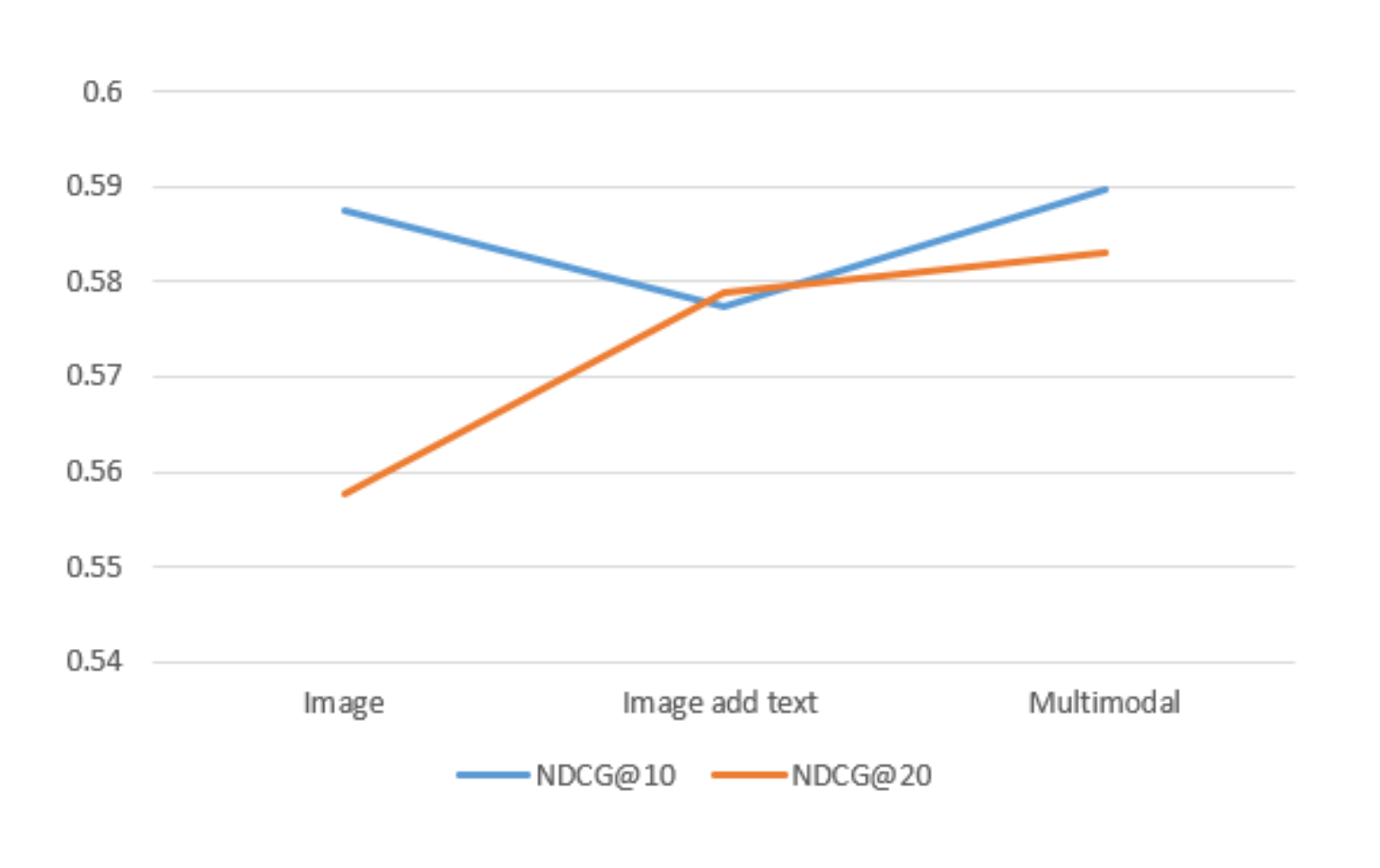}
      \label{fig_visualize_1_1}
    }
    \subfigure[HR@N on Movielens]{
      \includegraphics[width=0.23\textwidth]{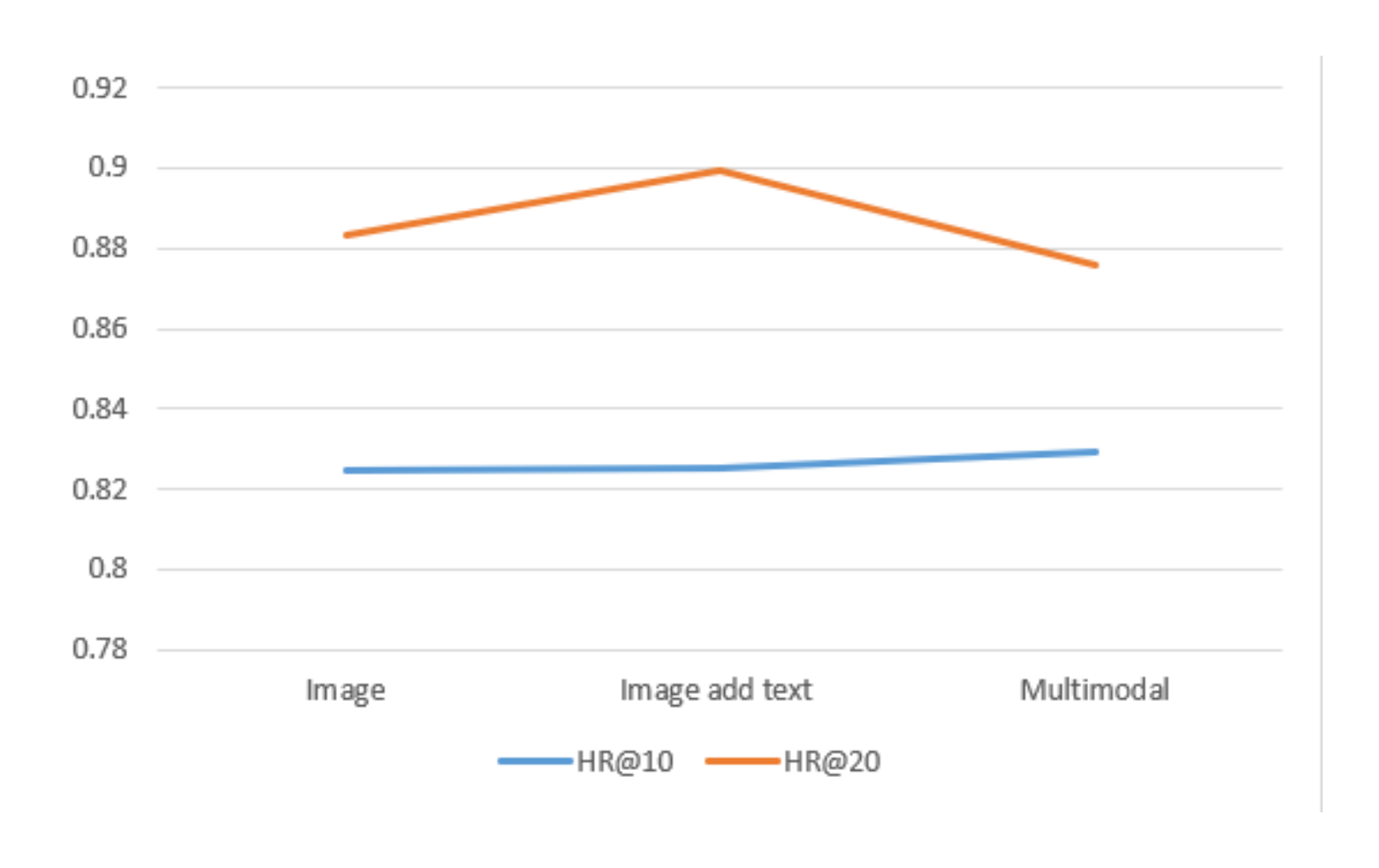}
      \label{fig_visualize_2_3}
    }
     \subfigure[NDCG@N on Amazon]{
      \includegraphics[width=0.23\textwidth]{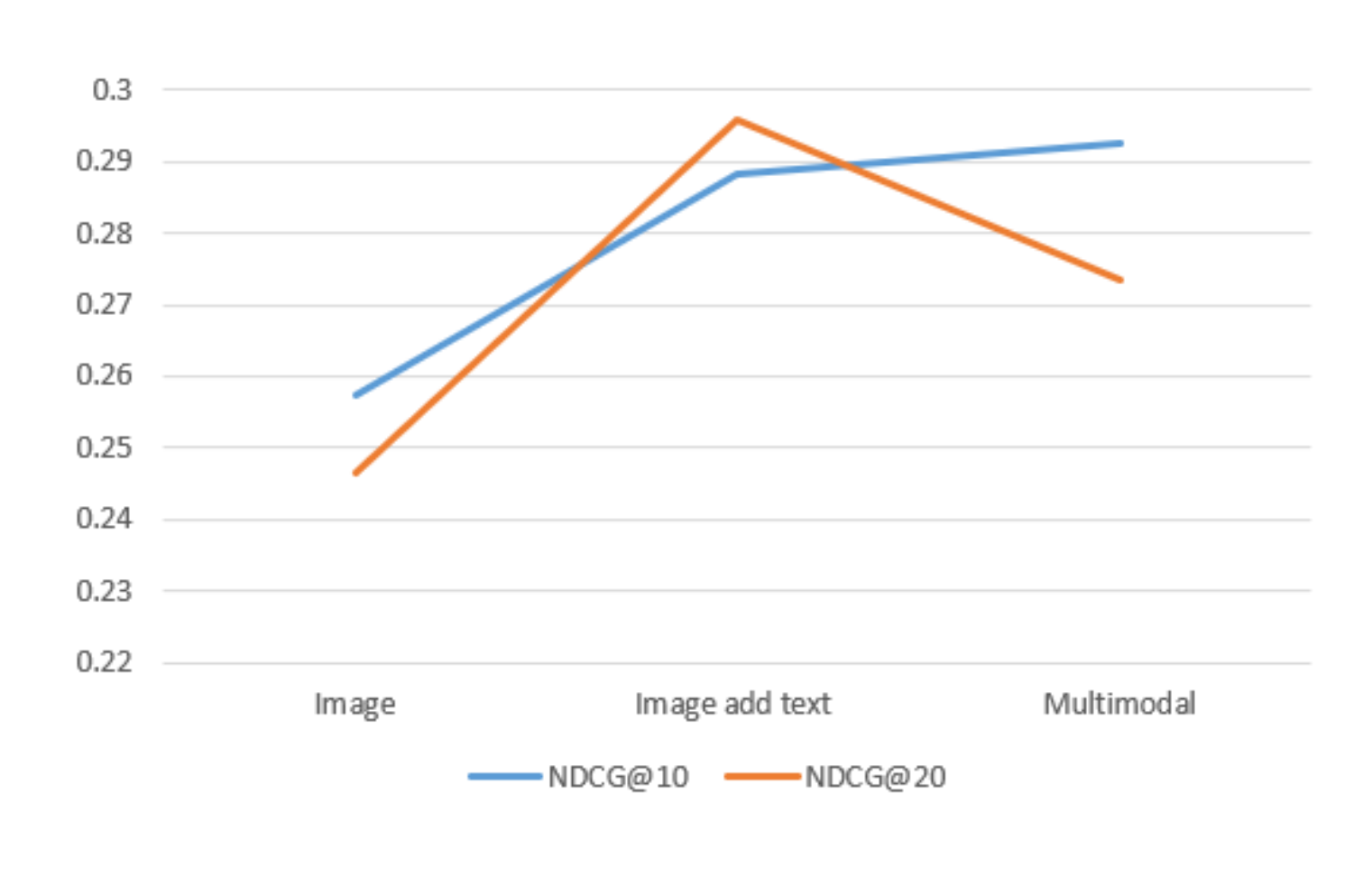}
      \label{fig_visualize_1_1}
    }
    \subfigure[HR@N on Amazon]{
      \includegraphics[width=0.23\textwidth]{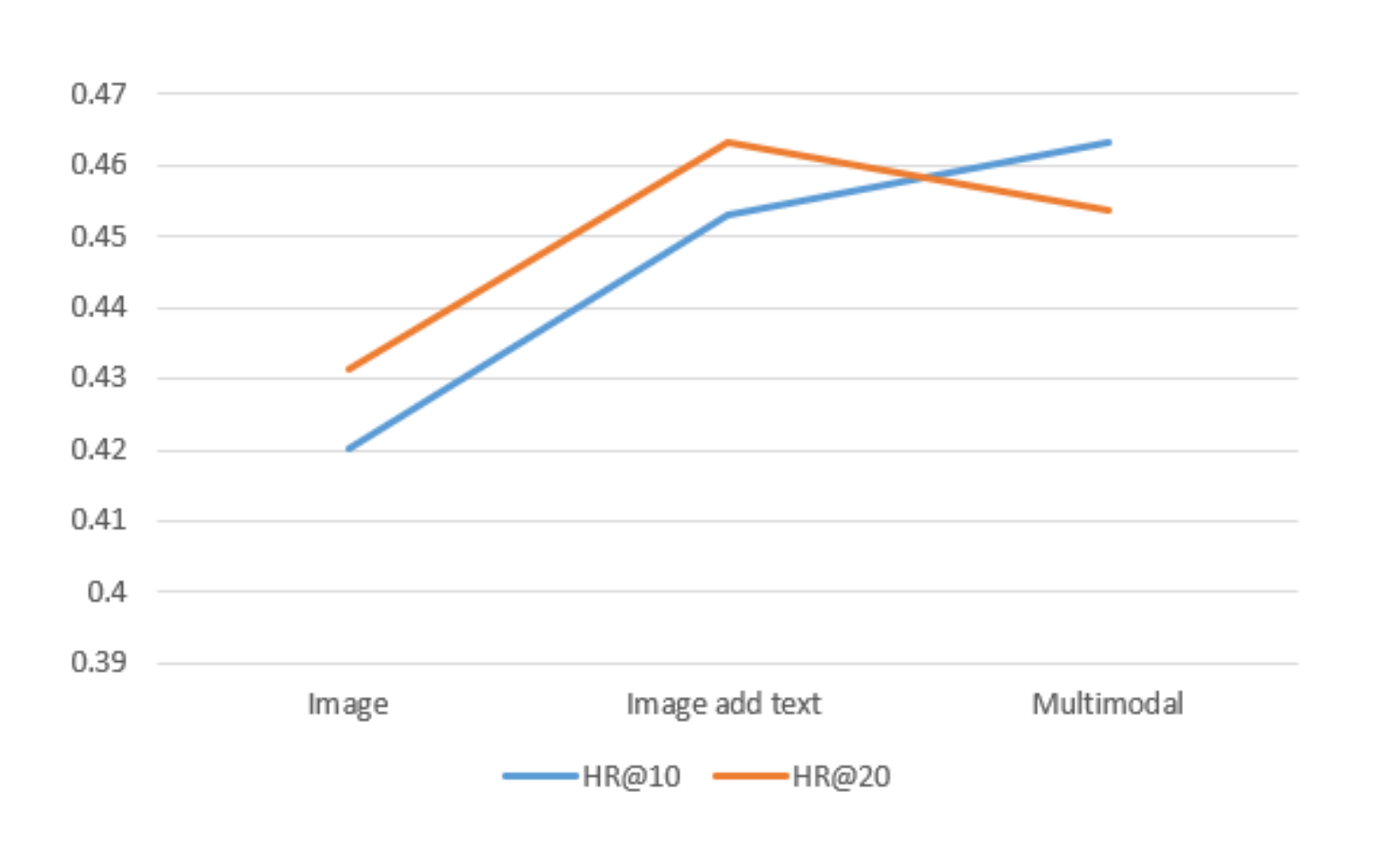}
      \label{fig_visualize_2_3}
    }
    \vspace{-10pt}
    \caption{Comparison of the results of the hidden factor-based model.}
    \label{fig_3}
    \vspace{-5pt}
 \end{figure} 

%% file: 5_conclusion.tex
\section{Conclusion}
There are still relatively few relevant studies conducted on micro-videos, and the main work of this thesis is to study and implement a personalized recommendation algorithm based on multimodal micro-video. The multimodal feature fusion multi-label classification model for micro-videos, which integrates subspace coding learning and multi-label relevance learning from multiple perspectives This model integrates subspace coding learning and multi-label relevance learning from multiple perspectives into a unified framework. The framework utilizes subspace coding networks to learn more representations of common representations across multiple views, while using graph convolutional networks to The framework uses a subspace coding network to learn more representations of common representations from multiple perspectives, and a graph convolutional network to mine semantic correlations between labels. The entire network is updated with a stochastic gradient descent-based alternating learning strategy for parameter solving. The interest-related product similarity model based on multimodal data sources is an end-to-end reordering model based on deep learning and attention mechanism for providing top-N recommendations. Finally, the multimodal data feature learning module, IRN and product similarity recommendation module are unified into one integrated model to achieve performance improvement and adapt to the addition or absence of different modal data. The model takes into account the multimodal data that people may pay more attention to when choosing products, thus greatly improving the accuracy and interpretability of the top-N recommendation task. Multimodality on the publicly Experimental results on the open micro-video dataset demonstrate the effectiveness of this method. 